# CLOSED EXPRESSIONS FOR LIE ALGEBRA INVARIANTS AND FINITE TRANSFORMATIONS

R. Aldrovandi[1,2], A.L. Barbosa[1,2] and L.P. Freitas[1,2]

Address: R. Aldrovandi, Rua Pamplona, 145  01405-900, São Paulo  SP  Brazil. E-Mail: RA @ AXP.IFT.UNESP.BR. phone (55) (11) 251 51 55; fax  (55) (11) 288 82 24.

---

[1] Instituto de Fisica Teorica, State University of Sao Paulo, São Paulo SP Brazil.
[2] Fellow of CNPq, Brasilia.

# ABSTRACT

A simple procedure to obtain complete, closed expressions for Lie algebra invariants is presented. The invariants are ultimately polynomials in the group parameters. The construction of finite group elements require the use of projectors, whose coefficients are invariant polynomials. The detailed general forms of these projectors are given. Closed expressions for finite Lorentz transformations, both homogeneous and inhomogeneous, as well as for Galilei transformations, are found as examples.

## 1. INTRODUCTION

Our objective here is twofold: first, to describe a general method to obtain in a simple way complete expressions for the elements of general Lie groups. Second, to present a treatment leading to complete, closed expressions for the Lie algebra invariants of matrix groups. In these times of increasing algebraic computing resources, it is always interesting to have such formulae as inputs for applications. Besides their obvious interest by themselves, the detailed forms of the invariants serve as a powerful checking control in the rather messy calculations leading to the expressions for finite transformations. These expressions, of course, are of interest by themselves. Even if particular cases (such as a rotation around the fixed axis Oz or a boost in the x-t plane for the Lorentz group) are enough for most purposes, only the general expressions show the complete interplay between the different components at work in a transformation. The procedure to get at them provide, furthermore, good illustrations for the results on the invariants. The two things help each other, and are better presented together.

The transformation parameters (rotation angles, boost rapidities, translations in time and space) are the components of the generic member of the Lie algebra written in a matrix basis formed by the generators. A finite transformation is a group element, the exponential of the algebra member. General expressions for group elements can in principle be obtained by simple order-by-order exponentiation, but for large matrices the identification of the successive powers can become very difficult. However, if we use the well-known definition of function of a matrix, only a few powers are necessary even in the general case.

The closed expressions for the Lie algebra invariants, scalar and operatorial, are presented in section 2. Though not quite necessary to arrive at the final exponentials for the group elements, they are interesting for checking purposes and become more and more useful as the matrix dimensions grow up. The treatment involving the characteristic polynomial and symmetric functions of its roots sheds light on many aspects of the question. In the later sections we illustrate the approach with the general expressions for Lorentz and Poincaré transformations. The method is particularly simple when all the roots (the matrix eigenvalues) are simple. The presence of multiple roots, as is the case for the Galilei group, requires a special treatment.

In section 3 we present the general method for obtaining the group elements. We called it the "Z-method", since it uses the "eigenprojectors" $Z_j = |\lambda_j\rangle\langle\lambda_j|$ of a finite matrix A with eigenvalues $\{\lambda_i\}$.

In section 4 we present an example of invariants which are very important in gauge theories: the closed differential forms describing the characteristic classes of fiber bundles on even-dimensional differentiable manifolds.

The general expression for a homogeneous Lorentz transformation is obtained in section 5, and that for an inhomogeneous transformation in section 6. For these examples we can use a simple version of the Z-method, which supposes all the eigenvalues to be distinct. The generic member of the Lie algebra of the Galilei group has a multiple eigenvalue, so that the procedure would not work. In section 7 we show, however, how to obtain it from a more general theory.

The appendix presents some results on symmetric functions which are largely used in section 2 and 4.

## 2. LIE ALGEBRA INVARIANTS

We start by recalling some well-known definitions. The characteristic polynomial (in the complex variable $\lambda$) of a matrix A is the determinant

$$\Delta(\lambda) = \det[\lambda I - A]. \qquad (2.1)$$

The roots of $\Delta(\lambda)$, solutions of the secular equation

$$\Delta(\lambda) = (\lambda-\lambda_1)(\lambda-\lambda_2)(\lambda-\lambda_3)\ldots(\lambda-\lambda_N) = 0, \qquad (2.2)$$

constitute the spectrum of A, the set Sp A = $\{\lambda_1, \lambda_2, \lambda_3, \ldots \lambda_N\}$ of complex numbers for which the resolvent, or characteristic matrix $[\lambda I - A]$, is not invertible.

We are interested in matrices A belonging to a Lie algebra. Clearly the trace of A is an invariant under the similarity transformation $A' = g^{-1}Ag$, as $\text{tr}(g^{-1}Ag) = \text{tr}(gg^{-1}A) = \text{tr } A$. The trace of any power of A will be invariant, as $\mathrm{tr}A'^k = \mathrm{tr}(g^{-1}Agg^{-1}Agg^{-1}Ag \ldots g^{-1}Ag) = \mathrm{tr}(g^{-1}A^k g) = \mathrm{tr}A^k$. The characteristic polynomial is an invariant because, once written in the canonical form

$$\Delta(\lambda) = \sum_{j=0}^{N} \lambda^{N-j} \varphi_j[A], \tag{2.3}$$

each coefficient $\varphi_j[A]$ is an invariant. These coefficients are invariant because they can be written in terms of powers of the above traces of powers of A. In order to see it, we need their detailed form, which involves Bell polynomials. Let us recall briefly what such polynomials are.

Given a formal series of the form $g(t) = \sum_{j=1}^{\infty} \frac{g_j}{j!} t^j$, the corresponding Bell polynomials are defined by

$$B_{nk}(g_1, g_2, \ldots, g_{n-k+1}) = \frac{1}{k!} \left\{ \frac{d^n}{dt^n} [g(t)]^k \right\}_{t=0}. \tag{2.4}$$

They are multivariable polynomials in the Taylor coefficients $g_i$. Their detailed expression is rather involved (Comtet, 1974), but they appear in the multinomial theorem,

$$\frac{1}{k!} \left( \sum_{j=1}^{\infty} \frac{g_j}{j!} t^j \right)^k = \sum_{n=k}^{\infty} \frac{t^n}{n!} B_{nk}(g_1, g_2, \ldots, g_{n-k+1}), \tag{2.5}$$

which provides the easiest way to obtain most of their properties. We use the alternative notations $B_{nk}[g] = B_{nk}(g_1, g_2, \ldots, g_{n-k+1}) = B_{nk}\{g_r\}$, the expression inside the latter bracket being the typical argument. By convention we put $B_{j0} = \delta_{j0}$. The matrices $B[g] = (B_{nk}[g])$ have many important properties (Aldrovandi and Monte Lima, 1980, 1983), but we shall only state those of interest to our particular case. A property of consequence in what follows is

$$B_{nk}(ag_1, a^2 g_2, \ldots, a^{n-k+1} g_{n-k+1}) = B_{nk}\{a^j g_j\} = a^n B_{nk}\{g_j\} \tag{2.6}$$

for a constant a. It comes from a simple change of variable $t \to at$ in the multinomial theorem. In particular, when a is only a sign, $a = -1$, we have

$$B_{nk}\{(-)^j g_j\} = (-)^n B_{nk}\{g_j\}. \tag{2.7}$$

Texts on Combinatorics in general call the summation

$$\frac{1}{N!}\sum_{m=0}^{N} B_{Nm}[g] \tag{2.8}$$

the "complete Bell polynomial" of g(x). We shall obtain the Lie algebra invariants precisely as complete Bell polynomials. The relation of the characteristic polynomial to the invariants is contained in the general formula

$$\det[\lambda I+zA] = \sum_{j=0}^{N} \frac{z^j \lambda^{N-j}}{j!} \sum_{m=0}^{j} B_{jm}\{(-)^{k-1}(k-1)!\mathrm{tr}(A^k)\}, \tag{2.9}$$

which is obtained by using $\ln(1 + x) = \sum_{j=1}^{\infty} \frac{(-)^{j-1}}{j} x^j$ in the formal identity $\det[\lambda I+zA] = \exp[\mathrm{tr}\{\ln[\lambda I+zA]\}]$ and then expanding. From this formula we recognize the coefficients of the characteristic polynomial (2.3) as

$$\varphi_j[A] = \frac{(-1)^j}{j!} \sum_{m=0}^{j} B_{jm}\{(-)^{k-1}(k-1)!\ \mathrm{tr}(A^k)\}. \tag{2.10}$$

These complete Bell polynomials contain only powers of traces of powers of A and are clearly invariant. A particularly beautiful relation comes out when we put $\lambda = 0$, $z = 1$ in (2.9):

$$\det A = \frac{1}{N!}\sum_{m=0}^{N} B_{Nm}\{(-)^{k-1}(k-1)!\mathrm{tr}(A^k)\}. \tag{2.11}$$

In the present case, this states the well-known fact that the zeroth-order coefficient of the characteristic polynomial of A is just the determinant of A.

Take now for the matrix a generic member

$$A = \omega^a J_a \tag{2.12}$$

of a Lie algebra of a Lie group with generators $J_a$, $a = 1, 2, \ldots, N$. The $\omega^a$'s are the transformation parameters. A group element will be of the form $g = \exp[A]$. The group acts on A by the adjoint N×N representation, that is, by similarity transformations: $A \to A' = g^{-1}Ag$. As the characteristic polynomial does not change under these transformations, it is an example of an invariant polynomial on a Lie algebra. Summing up, $\Delta(\lambda)$ is a polynomial in the variable $\lambda$ whose coefficients are functions $\varphi_j(\omega) = \varphi_j[A]$ of the parameters, which are themselves Bell polynomials in the traces of powers of A. The rank of the Lie algebra is the number of coefficients $\varphi_j(\omega)$ which are functionally independent, that is, the rank of the matrix $(\partial \varphi_j / \partial \omega^i)$. Some examples of these invariant polynomials are:

$\varphi_o(\omega) = 1;$
$\varphi_1(\omega) = -\operatorname{tr} A;$
$\varphi_2(\omega) = \frac{1}{2} [-\operatorname{tr} A^2 + (\operatorname{tr} A)^2];$
$\varphi_3(\omega) = -\frac{1}{3} \operatorname{tr} A^3 + \frac{1}{2} (\operatorname{tr} A)(\operatorname{tr} A^2) - \frac{1}{6} (\operatorname{tr} A)^3;$
$\varphi_4(\omega) = -\frac{1}{4} \operatorname{tr} A^4 + \frac{1}{3}(\operatorname{tr} A)(\operatorname{tr} A^3) + \frac{1}{8} (\operatorname{tr} A^2)^2 - \frac{1}{4} (\operatorname{tr} A)^2(\operatorname{tr} A^2) + \frac{1}{24} (\operatorname{tr} A)^4;$
. . .
$\varphi_N(\omega) = (-)^N \det A.$  (2.13)

These relations lead to the introduction of multilinear objects which are invariant symmetric tensors, such as

$$\eta^{(0)} = 1;\ \eta^{(1)}{}_a = -\operatorname{tr} J_a;\ \eta^{(2)}{}_{ab} = -\frac{1}{2} [\operatorname{tr}(J_a J_b) - \operatorname{tr}(J_a)\operatorname{tr}(J_b)];$$

$$\eta^{(3)}{}_{abc} = -\frac{1}{3} \left[\operatorname{tr}(J_a J_b J_c) - \frac{3}{2} (\operatorname{tr} J_a)(\operatorname{tr} J_b J_c) + \frac{1}{2} \operatorname{tr}(J_a)\operatorname{tr}(J_b)\operatorname{tr}(J_c)\right].$$

The general expression is

$$\eta^{(n)}{}_{a_1 a_2 \ldots a_n} = \frac{(-1)^n}{n!} \sum_{m=0}^{n} B_{nm}\{(-)^{k-1}(k-1)!\,\operatorname{tr}(J_{a_1} J_{a_2} \ldots J_{a_k})\}. \quad (2.14)$$

These tensors provide invariant objects by contraction,

$$T = T^{a_1 a_2 \ldots a_n} \eta^{(n)}{}_{a_1 a_2 \ldots a_n}, \quad (2.15)$$

and the invariant Casimir operators in the enveloping algebra:

$$C^{(n)} = \eta^{(n)}{}_{a_1 a_2 \ldots a_n} J^{a_1} J^{a_2} \ldots J^{a_n}. \quad (2.16)$$

The best known of these tensors is the Killing form $\gamma_{ab}$. It is usually defined, for the case in which $\operatorname{tr} J_a = 0$ for all the generators, as $\gamma_{ab} = -\frac{1}{2} \operatorname{tr} J_a J_b$ and leads to the Casimir operator $\gamma_{ab} J^a J^b$. It plays a fundamental role in the classification of Lie groups (Goldberg, 1962). For semisimple groups, $\gamma_{ab}$ is a metric (that is, a symmetric non-degenerate bilinear form), the Killing-Cartan metric. This form will be of definite sign for compact groups, non-definite for non-compact groups and degenerate for non-semisimple groups.

Of course, a function of invariants is also an invariant, and it is sometimes convenient to choose another set. Those given above are, however, because of their simple origin, the basic ones. The results may seem rather trivial when a diagonalized matrix is considered, but formulae (2.13) are independent of matrix basis. Some deep consequences can be found, related to the detailed algebraic structure of the Lie algebra. For example, a semisimple algebra is the direct product of simple sub-algebras and its characteristic polynomial is always of the form $\Delta(\lambda) = \prod_k \Delta_k(\lambda)$, where the $\Delta_k$'s are the characteristic polynomials of the simple subalgebras.

Orthogonal or pseudo-orthogonal groups are usually introduced as groups of real transformations which preserve a real bilinear non-singular form $\eta$. If g is a group element, this means that $g^T \eta g = \eta$, where $g^T$ is the transposed matrix. If $g = \exp[A]$, with A a member of the group Lie algebra, then this implies $A^T = -\eta A \eta^{-1}$ and the traces of the odd powers of A vanish. Actually, the odd coefficients in the characteristic polynomial (that is, the odd-order invariants) vanish. The argument is the following. Take $z = -1$ in (2.9),

$$\det[\lambda I - A] = \sum_{j=0}^{N} \frac{(-)^j \lambda^{N-j}}{j!} \sum_{m=0}^{j} B_{jm}\{(-)^{k-1}(k-1)! \operatorname{tr}(A^k)\} . \qquad (2.17)$$

The determinant for the transposed matrix must be the same: $\det[\lambda I - A] = \det[\lambda I - A^T]$. As $A^T = -\eta A \eta^{-1}$, a sign factor $(-)^k$ appears in the arguments of the Bell polynomials: $B_{jm}\{(-)^{k-1}(k-1)!\operatorname{tr}[(A^T)^k]\} = B_{jm}\{(-)^k(-)^{k-1}(k-1)!\operatorname{tr}(A^k)\} = (-)^j B_{jm}\{(-)^{k-1}(k-1)!\operatorname{tr}(A^k)\}$ by (2.7). Consequently, we have also

$$\det[\lambda I - A] = \sum_{j=0}^{N} \frac{\lambda^{N-j}}{j!} \sum_{m=0}^{j} B_{jm}\{(-)^{k-1}(k-1)!\operatorname{tr}(A^k)\}.$$

Comparison with (2.17) shows that the coefficients must vanish for odd values of j: $\varphi_{2p+1}(\omega) = 0$. The invariants of (2.13) reduce to simpler forms, like

$$\varphi_0(\omega) = 1;\ \varphi_1(\omega) = 0;\ \varphi_2(\omega) = -\tfrac{1}{2}\operatorname{tr}A^2;\ \varphi_3(\omega) = 0;$$

$$\varphi_4(\omega) = -\frac{1}{4} \operatorname{tr} A^4 + \frac{1}{8} (\operatorname{tr} A^2)^2; \ldots \tag{2.18}$$

For unitary groups written in terms of hermitian generators, $A^T = A^*$, the only conclusion is that the invariants are real, $\varphi_j(\omega)^* = \varphi_j(\omega)$. The extra condition $\det[g = \exp[iA]] = 1$ for special unitary groups leads, of course, to $\operatorname{tr} A = 0$.

The rank of the Lie algebra being the number of independent invariants, the higher the rank the more multilinear forms there will be. The Lorentz and the Poincaré groups are of rank 2 and have bilinear and quadrilinear forms. Another rank 2 case is the group SU(3) of special unitary 3×3 matrices, which has a bilinear and a trilinear invariant. Let us use this group to illustrate some of the above results. In terms of parameters $\alpha = \{\alpha_k, k = 1, 2, \ldots, 8\}$, and using the Gell-Mann basis $\{T_k\}$ for the algebra (Lee, 1981), the group element is

$$g[\alpha] = \exp[i\alpha_k T_k/2] . \tag{2.19}$$

The generic algebra element will be

$$W = \alpha_k T_k = \begin{pmatrix} \alpha_3 + 3^{-1/2}\alpha_8 & \alpha_1 - i\alpha_2 & \alpha_4 - i\alpha_5 \\ \alpha_1 + i\alpha_2 & -\alpha_3 + 3^{-1/2}\alpha_8 & \alpha_6 - i\alpha_7 \\ \alpha_4 + i\alpha_5 & \alpha_6 + i\alpha_7 & -2 \times 3^{-1/2}\alpha_8 \end{pmatrix} . \tag{2.20}$$

With the notation $\alpha_{12} = \alpha_1 + i\alpha_2$, $\alpha_{45} = \alpha_4 + i\alpha_5$, $\alpha_{67} = \alpha_6 + i\alpha_7$, $\alpha_{38} = \alpha_3 + 3^{-1/2}\alpha_8$, and $\alpha^*_{38} = \alpha_3 - 3^{-1/2}\alpha_8$, we obtain

$$W^2 = \begin{pmatrix} |\alpha_{12}|^2 + |\alpha_{45}|^2 + (\alpha_{38})^2 & \alpha^*_{45}\alpha_{67} + 2\times 3^{-1/2}\alpha^*_{12}\alpha_8 & \alpha^*_{12}\alpha^*_{67} + \alpha^*_{45}\alpha^*_{38} \\ \alpha_{45}\alpha^*_{67} + 2\times 3^{-1/2}\alpha_{12}\alpha_8 & |\alpha_{12}|^2 + |\alpha_{67}|^2 + (\alpha^*_{38})^2 & \alpha_{12}\alpha^*_{45} - \alpha^*_{67}\alpha_{38} \\ \alpha_{12}\alpha_{67} + \alpha^*_{45}\alpha_{38} & \alpha^*_{12}\alpha_{45} - \alpha_{67}\alpha_{38} & |\alpha_{45}|^2 + |\alpha_{67}|^2 + 4\alpha_8^2/3 \end{pmatrix} \tag{2.21}$$

As $\operatorname{tr} W = 0$, we expect the characteristic polynomial $\Delta(\lambda) = \lambda^3 \varphi_0 + \lambda^2 \varphi_1 + \lambda \varphi_2 + \varphi_3$ to have $\varphi_0 = 1$; $\varphi_1 = 0$; $\varphi_2 = -(1/2)\operatorname{tr} W^2$; $\varphi_3 = -\det W = -(1/3)\operatorname{tr} W^3$. We find indeed from (2.20) that the polynomial has the form

$$\Delta(\lambda) = \lambda^3 + \lambda \varphi_2 + \varphi_3 = \lambda^3 - \frac{\lambda}{2} \operatorname{tr} W^2 - \det W, \tag{2.22}$$

a result which can also be obtained by taking in (2.3) the coefficients (2.13). Applying the secular equation to the matrix, $W^3 - \frac{1}{2} W \, trW^2 - I \det W = 0$, and taking the trace, we find immediately that $trW^3 = 3 \det W$, a particular example of (2.11). We can, of course, obtain $W^3$ directly and check. It will be simpler in our case to use $W^3 = -(I \varphi_3 + W \varphi_2)$. The first invariant is that related to the Cartan metric on the 8-dimensional SU(3) manifold,

$$\varphi_2 = -\frac{1}{2} \, trW^2 = -\sum_{i=1}^{8} \alpha_i^2 = -\gamma_{ij}\alpha_i\alpha_j, \qquad (2.23)$$

with $\gamma_{ij} = \delta_{ij}$ chosen conventionally as positive and Einstein summation notation adopted in the last expression. The corresponding Casimir invariant is the operator $C^{(2)} = -\sum_{i=1}^{8} T_i^2$. The other invariant is not so simple, as it relates to a trilinear form $\eta_{ijk}$ defined through

$$\varphi_3 = -\frac{1}{3} \, tr \, W^3 = \frac{1}{3} \sum_{i,j,k=1}^{8} \alpha_i\alpha_j\alpha_k \, tr \, (T_iT_jT_k) = -\sum_{i,j,k=1}^{8} \eta_{ijk} \, \alpha_i\alpha_j\alpha_k, \qquad (2.24)$$

corresponding to the Casimir $C^{(3)} = -\sum_{i,j,k=1}^{8} \eta_{ijk}T_iT_jT_k$. The components $\eta_{ijk}$ can be picked up from the explicit expression of det W,

$$\eta_{ijk} \, \alpha_i\alpha_j\alpha_k = 2 \, (\alpha_1\alpha_4\alpha_6 + \alpha_2\alpha_5\alpha_6 - \alpha_2\alpha_4\alpha_7 + \alpha_1\alpha_5\alpha_7) + \alpha_3\alpha_4^2 + \alpha_3\alpha_5^2 - \alpha_3\alpha_6^2 -$$
$$\alpha_3\alpha_7^2 + 3^{-1/2}[2(\alpha_1^2 + \alpha_2^2 + \alpha_3^2) - \alpha_4^2 - \alpha_5^2 - \alpha_6^2 - \alpha_7^2 - 2\,\alpha_8^2/3]\alpha_8. \qquad (2.25)$$

## 3. FUNCTIONS OF MATRICES AND THE Z-METHOD

Let us begin by saying a few words on functions of matrices in general (Gantmacher, 1990). Suppose a complex function $F(\lambda)$ is given which can be expanded as a power series $F(\lambda) = \sum_{k=0}^{\infty} c_k(\lambda - \lambda_0)^k$ on some convergence disk $|\lambda - \lambda_0| < r$. Then the function $F(A)$, now with the matrix A in the argument, is defined as the matrix $F(A) = \sum_{k=0}^{\infty} c_k (A - \lambda_0)^k$ and is meaningful whenever the eigenvalues of A lie within the convergence circle. In particular, the exponential is always well defined.

Given an N×N matrix A of eigenvalues $\lambda_1, \lambda_2, \ldots, \lambda_N$, the set of eigenprojectors $\{Z_j[A] = |\lambda_j\rangle\langle\lambda_j|\}$ constitutes a basis, in which A is written as

$$A = \sum_{j=1}^{N} \lambda_j Z_j[A] . \qquad (3.1)$$

Then the function F(A) can be equivalently defined as the matrix

$$F(A) = \sum_{j=1}^{N} F(\lambda_j) Z_j[A] . \qquad (3.2)$$

In particular, we shall be looking for

$$e^A = \sum_{j=1}^{N} e^{\lambda_j} Z_j[A] . \qquad (3.3)$$

The $Z_j$'s have remarkable properties. Besides being projectors (that is, idempotents, $Z^2_j = Z_j$), they can be normalized so that tr $(Z_j) = 1$ for each j (for a general idempotent Z, tr $Z$ = rank Z is the dimension of the subspace into which it projects). They are then orthonormal by the trace, $\text{tr}(Z_i Z_j) = \delta_{ij}$. The necessary results $\text{tr}[F(A)] = \sum_{j=1}^{N} F(\lambda_j)$, $\text{tr}[A^k Z_j] = (\lambda_j)^k$, etc., follow immediately. If A is a normal matrix diagonalized by a matrix U, $UAU^{-1} = A_{\text{diagonal}}$, the entries of $Z_k$ are $(Z_k)_{rs} = U^{-1}_{rk} U_{ks}$ (fixed "k"). Finding U is equivalent to finding the projectors. For a N×N matrix, a limited number of powers is enough to determine the basis $\{Z_j[A]\}$: writing (3.2) for the power functions $F(A) = A^0 = I, A^1, A^2, \ldots, A^{N-1}$, we have

$$I = \sum_{j=1}^{N} Z_j; \quad A = \sum_{j=1}^{N} \lambda_j Z_j; \quad A^2 = \sum_{j=1}^{N} \lambda_j^2 Z_j; \quad \ldots ; \quad A^k = \sum_{j=1}^{N} \lambda_j^k Z_j;$$

$$\ldots ; A^{N-1} = \sum_{j=1}^{N} \lambda_j^{N-1} Z_j. \tag{3.4}$$

For $k \geq N$, the $A^k$'s are no longer independent. Inversion of the above expressions leads to the closed forms

$$Z_j[A] = \frac{(A - \lambda_1)(A - \lambda_2)\ldots(A - \lambda_{j-1})(A - \lambda_{j+1})\ldots(A - \lambda_{N-1})(A - \lambda_N)}{(\lambda_j - \lambda_1)(\lambda_j - \lambda_2)\ldots(\lambda_j - \lambda_{j-1})(\lambda_j - \lambda_{j+1})\ldots(\lambda_j - \lambda_{N-1})(\lambda_j - \lambda_N)}. \tag{3.5}$$

This is, of course, a polynomial in A, with coefficients dependent on the eingenvalues. Here all the eigenvalues must be distinct. Equation (3.2) becomes

$$F(A) = \sum_j \left\{ \prod_{k \neq j} \frac{A - \lambda_k}{\lambda_j - \lambda_k} \right\} F(\lambda_j). \tag{3.6}$$

This expression is equivalent to an elegant equation, stating the vanishing of a formal determinant:

$$\begin{vmatrix} F(A) & F(\lambda_1) & F(\lambda_2) & F(\lambda_3) & \ldots & F(\lambda_N) \\ I & 1 & 1 & 1 & \ldots & 1 \\ A & \lambda_1 & \lambda_2 & \lambda_3 & \ldots & \lambda_N \\ A^2 & \lambda_1^2 & \lambda_2^2 & \lambda_3^2 & \ldots & \lambda_N^2 \\ A^3 & \lambda_1^3 & \lambda_2^3 & \lambda_3^3 & \ldots & \lambda_N^3 \\ \cdot & \cdot & \cdot & \cdot & \ldots & \cdot \\ \cdot & \cdot & \cdot & \cdot & \ldots & \cdot \\ A^{N-1} & \lambda_1^{N-1} & \lambda_2^{N-1} & \lambda_3^{N-1} & \ldots & \lambda_N^{N-1} \end{vmatrix} = 0. \tag{3.7}$$

Thus, in order to obtain $e^A = \sum_j \{\prod_{k \neq j} \frac{A - \lambda_k}{\lambda_j - \lambda_k}\} e^{\lambda_j}$, it is necessary to find (i) the eigenvalues of A and (ii) the detailed form of the first (N-1) powers of A. What we have done is to start from a basis $\{A^k\}$ for the functions of A, and go to the projector basis. Actually, any set of independent polynomials in A could be used as the starting basis. Their use can prove necessary to avoid that the above determinant vanish identically. Thus, once we know the projectors, we can use them as the basis polynomials in A and the above determinant becomes

$$\begin{vmatrix} F(A) & F(\lambda_1) & F(\lambda_2) & \ldots & F(\lambda_N) \\ Z_1[A] & 1 & 0 & \ldots & 0 \\ Z_2[A] & 0 & 1 & 0.. & 0 \\ \cdot & \cdot & \cdot & \ldots & \cdot \\ Z_N[A] & 0 & 0 & \ldots & 1 \end{vmatrix} = 0, \qquad (3.8)$$

which is clearly the same as (3.2).

To obtain the closed expressions of the projectors in terms of powers of A and the invariants, we need some results on symmetric functions, of which a brief account is given in the appendix. The expression (3.5) is actually the ratio of two generating functions of the so-called elementary symmetric functions of the "alphabet" $\{\lambda_1, \lambda_2, \lambda_3, \ldots, \lambda_N\}$ with one missing letter. The detailed calculation is also shown in the appendix and gives each projector (3.5) in terms of the invariants (2.13):

$$Z_i[A] = \sum_{k=0}^{N} \frac{\sum_{j=0}^{k} \lambda_i^{j-k} \varphi_{N-j}[A]}{\sum_{n=0}^{N} \sum_{j=0}^{n} \lambda_i^{j} \varphi_{N-j}[A]} A^k. \qquad (3.9)$$

The reader may have qualms about the case where some root $\lambda_0 = 0$. The good rule is to perform first the computation of the projectors with the symbols $\lambda_k$, without putting in the values, and then multiply numerator and denominator by $\det A = \Pi_k \lambda_k$. This procedure will dispose of any problem with eventual vanishing eigenvalues. The closed expressions (3.9) are more useful for larger matrices. For small ones they can, however, be used to check the results obtained by direct computation. We have found them very useful while checking the calculations presented below with the Mathematica™ software system.

When the matrix A is a member of a Lie algebra, we have a curious problem, touching Galois theory. As we saw in the section 2, the coefficients of the characteristic polynomial of a matrix A are written in terms of invariant polynomials and the roots $\lambda_i$. Suppose we are able to solve the characteristic polynomial. This means that we can write the $\lambda_i$'s in terms of the polynomials coefficients, that is, of the invariants. In this case, the coefficient of each $A^k$ in (3.9) is an invariant. Notice, however, our basic assumption: we should be able to solve the polynomial. Fortunately, this will be the case for the Lie algebras of groups of physical interest examined in the following. This is not the case in general for N ≥ 5.

In plain sight expressions like (3.5) become undefined when two or more eigenvalues coincide. This formulation works only for nondegenerate matrices. The method will consequently only apply for Lie algebras whose generic elements have no degenerate eigenvalues. This happens for the SU(3), Lorentz and Poincaré cases, but not for the Galilei generic algebra element.

The general expressions in the presence of degenerate eigenvalues are rather involved. We shall only present a mnemonic rule, addressing the more interested reader to a complete reference (Gantmacher, 1990). For that it is better to look back at (3.7). Suppose that the eigenvalues are degenerate: that $n_1$ of the roots are equal to $\lambda_1$, $n_2$ of the roots are $\lambda_{n_1+1}$, and so on. The determinant becomes, of course, identically null, and the equation gives no real information. We have to correct it. The rule is then the following: (i) the first column remains as it is; (ii) the second, headed by $F(\lambda_1)$, also; (iii) the next $(n_1-1)$ columns are replaced by the successive derivatives of order 1, 2, . . . , $(n_1-1)$ of $F(\lambda_1)$; (iv) then comes the column headed by $F(\lambda_{n_1+1})$; (v) the next $(n_2-1)$ columns are the derivatives of that column, up to that headed by $F^{(n_1+n_2-1)}(\lambda_{n_1+1})$; and so on for all the columns. Notice that we first derive the formal expressions and only after put in the exact values of the roots. The general aspect of the formal determinantal equation becomes

$$\begin{vmatrix} F(A) & F(\lambda_1) & F'(\lambda_1) & \ldots F^{(n_1-1)}(\lambda_1) & F(\lambda_2) & \ldots F^{(n_2-1)}(\lambda_2) & \ldots F(\lambda_n) & \ldots F^{(n_n-1)}(\lambda_n) \\ I & 1 & 1 & \ldots 1 & 1 & \ldots 1 & \ldots 1 & \ldots 1 \\ A & \lambda_1 & 1 & \ldots 0 & \lambda_2 & \ldots 0 & \ldots \lambda_n & \ldots \\ A^2 & \lambda_1^2 & 2\lambda_1 & \ldots & \lambda_2^2 & \ldots & \ldots \lambda_n^2 & \ldots \\ A^3 & \lambda_1^3 & 3\lambda_1^2 & \ldots & \lambda_2^3 & \ldots & \ldots \lambda_n^3 & \ldots \\ \ldots & \ldots & \ldots & \ldots & & & & \\ \ldots & \ldots & \ldots & \ldots & & & & \\ \ldots & \ldots & \ldots & \ldots & & & & \\ \ldots & \ldots & \ldots & \ldots & & & & \\ \ldots & \ldots & \ldots & \ldots & & & & \\ \ldots & \ldots & \ldots & \ldots & & & & \\ A^{N-1} & \lambda_1^{N-1} & (N-1)\lambda_1^{N-2} & \ldots & & & & \end{vmatrix} = 0.$$

(3.10)

We shall illustrate it later with the Galilei group. Also here it may happen that, for a certain choice of the monomials in the first column, the determinant vanishes identically. In this case, we must replace the monomials by another collection (eventually of polynomials), in order to have a nontrivial equation.

## 4. CHARACTERISTIC CLASSES

Perhaps the most beautiful examples of the invariants described above are the differential forms describing the characteristic classes of fiber bundles on even-dimensional differentiable manifolds (Kobayashi and Nomizu 1963). These eminently mathematical objects are of great interest in gauge theories and have received a lot of attention of physicists (Eguchi, Gilkey and Hanson 1980; Zumino 1985). We shall only consider briefly their formal aspects, and confine ourselves to Chern classes. There are actually two interrelated kinds of classes going under Chern's name: the Chern classes proper and the Chern characters (Nakahara 1990). We shall find easily the closed, complete expressions of their interrelations. Consider the complex linear group GL(N,**C**) of all invertible N×N matrices with complex entries and let A be an element of its Lie algebra G'L(N,**C**). To conform to current mathematical notation it is enough to take formula (2.9) with $z = \frac{i}{2\pi}$, so that now the invariant polynomial functions are

$$\varphi_k[A] = (\frac{i}{2\pi})^k \frac{1}{k!} \sum_{m=0}^{k} B_{km}\{(-)^{r-1}(r-1)!tr(A^r)\}. \tag{4.1}$$

Given a complex vector bundle E over a manifold M with typical fiber $\mathbf{C}^N$ and associated to a principal bundle P, the k-th Chern class $c_k(E)$ will be characterized by a closed differential form $c_k$ of degree 2k on M. This form is that (unique) form whose pull-back is $\varphi_k(F)$, where $F = d\Gamma + \Gamma_H\Gamma$ is the curvature of a connection $\Gamma$ on P. After they are found, the cohomology classes represented by the $c_k$'s are shown to be independent of $\Gamma$ and, consequently, related to the more basic, topological features of the bundle. If $\pi: P \rightarrow M$ is the bundle projection and $\pi^*$ the corresponding pull-back,

$$\pi^*(c_k) = \varphi_k(F) = \frac{i^k}{(2\pi)^k k!} \sum_{m=0}^{k} B_{km}\{(-)^{r-1}(r-1)! \operatorname{tr}[(F)^r]\} = \sigma_k[iF/2\pi], \quad (4.2)$$

the symmetric function of the alphabet whose letters are the eigenvalues of the matrix $(iF/2\pi)$. Then,

$$c_0(E) = 1;$$
$$c_1(E) = \frac{i}{2\pi} \operatorname{tr} F\,;$$
$$c_2(E) = \frac{1}{8\pi^2}[\operatorname{tr}F^2 - (\operatorname{tr}F)^2];$$
$$c_3(E) = -\frac{i}{8\pi^3}\left[\frac{1}{3}\operatorname{tr}F^3 + \frac{1}{2}(\operatorname{tr}F)(\operatorname{tr}F^2) - \frac{1}{6}(\operatorname{tr}F)^3\right];$$
$$\ldots$$
$$c_N(E) = \left(\frac{i}{2\pi}\right)^N \det F. \quad (4.3)$$

Notice that here N is the fiber dimension. These relations are simple adaptations of (2.13). For orthogonal groups, the first invariants reduce to the simpler forms of (2.18),

$$c_0(E) = 1;\ c_1(E) = 0;\ c_2(E) = \frac{1}{8\pi^2}\operatorname{tr}F^2;\ c_3(E) = 0;$$
$$c_4(E) = -\frac{1}{64\pi^4}\operatorname{tr}F^4 + \frac{1}{128\pi^4}(\operatorname{tr}F^2)^2;\ \ldots \quad (4.4)$$

Handling matrices of forms in general requires some attention, because the non-commutativity of forms is added to that of matrices. Here only 2-forms appear, which are commutative. Of course, the higher-order classes are of limited physical interest, as products of forms vanish when the resulting degree surpasses the space dimension. On 4-dimensional spacetime, only invariant polynomials up to $\varphi_4$ are non-vanishing.

The Chern total character is defined as

$$\operatorname{ch}(E) = \operatorname{tr}\left(\exp\left[\frac{i}{2\pi}F\right]\right) = \sum_{k=0}^{\infty} \frac{i^k}{(2\pi)^k k!} \operatorname{tr}F^k = \sum_{k=0}^{N} \frac{1}{k!} s_k[iF/2\pi], \quad (4.5)$$

where the $s_k$'s are the power symmetric functions defined in appendix A, with the additional conventions $s_{k>N}[iF/2\pi] = 0$ and $s_0[iF/2\pi] = N$. The summand

$$\text{ch}_k(E) = \frac{i^k}{(2\pi)^k k!} \text{tr} F^k = \frac{1}{k!} s_k[iF/2\pi] \tag{4.6}$$

is the k-th Chern character. The relations between classes and characters are, for $j,k \geq 0$, mere reflections of (A.6) and (A.7):

$$c_j(E) = \frac{1}{j!} \sum_{m=0}^{j} B_{jm} \{(-)^{k-1} (k-1)! \, k! \, \text{ch}_k(E)\}; \tag{4.7}$$

$$k! \, \text{ch}_k(E) = \frac{(-1)^{k-1}}{(k-1)!} \sum_{j=0}^{k} (-)^{j-1}(j-1)! \, B_{kj} \{r! c_r(E)\}. \tag{4.8}$$

Our concern here is simply to exhibit these closed expressions for the relationships. We have only considered the complex linear groups. The interested reader will find in the quoted references details on Pontryagin classes $g_{2k}(F) = (-)^k f_{2k}(F)$, Euler classes, etc.

The relations (4.3) are, as said, simple adaptations of (2.13) and lead to the introduction of multilinear objects like (2.14), which are invariant tensors and allow to obtain more general invariants. The usual Lagrangian density for gauge fields, for example, is the invariant $\text{tr} \, F\tilde{F} = \gamma_{ab} F^a \tilde{F}^b$, where $\gamma_{ab}$ is the Killing form and $\tilde{F}$ is the dual of F in Minkowski space. The corresponding Chern class, $c_2(E)$, would be a "topological Lagrangian density". Current gauge theories use groups for which $\text{tr} \, F = 0$, so that only the term in $\text{tr} \, F^2$ appear in the Lagrangian. A gauge theory for the linear group GL(4, R), for example, could have (Aldrovandi and Stedile, 1984) an extra term in $(\text{tr} F)(\text{tr} \, \tilde{F})$.

We now apply the Z-method to some cases of special importance. We shall, as a rule, use anti-hermitian generators in what follows.

## 5. LORENTZ TRANSFORMATIONS

The Lorentz generators can be taken as the 4×4 matrices $J_{\alpha\beta}$ with entries $(J_{\alpha\beta})\gamma\delta = \eta_{\alpha\gamma}\eta_{\beta\delta} - \eta_{\beta\gamma}\eta_{\alpha\delta}$, with $\alpha, \beta, \gamma, \delta = 0, 1, 2, 3$ and where $\eta = \text{diag}(1,-1,-1,-1)$ is the Lorentz metric, responsible for the lowering and raising of indices in all the formulae below. They will satisfy

$$[J^{\alpha\beta}, J^{\gamma\delta}] = \eta^{\beta\gamma}J^{\alpha\delta} + \eta^{\alpha\delta}J^{\beta\gamma} - \eta^{\beta\delta}J^{\alpha\gamma} - \eta^{\alpha\gamma}J^{\beta\delta}. \tag{5.1}$$

A general algebra element will be

$$\mathbf{A} = \tfrac{1}{2}\omega_{\alpha\beta}\, J^{\alpha\beta}. \tag{5.2}$$

The double-index notation (with Einstein convention) we are adopting avoids the use of too large matrices and the numerical factor accounts for double-counting. The parameters $\omega_{\alpha\beta}$ will be the rotation group angles $\boldsymbol{\omega} = (\omega_1, \omega_2, \omega_3)$, with $\omega_{ij} = \varepsilon_{ijk}\omega^k$ and (with $\boldsymbol{\beta} = \mathbf{v}/c$, $\beta = |\boldsymbol{\beta}|$) the imaginary angles $\omega_{0j} = \zeta_j$, actually the boost rapidities collected in the vector $\boldsymbol{\zeta} = (\zeta_1, \zeta_2, \zeta_3) = \hat{\boldsymbol{\beta}}\,\tanh^{-1}[\beta] = \mathbf{v}/|\mathbf{v}|\,\tanh^{-1}[|\mathbf{v}/c|]$. The algebra element becomes

$$\mathbf{A} = \begin{pmatrix} 0 & -\zeta^1 & -\zeta^2 & -\zeta^3 \\ -\zeta^1 & 0 & -\omega^3 & \omega^2 \\ -\zeta^2 & \omega^3 & 0 & -\omega^1 \\ -\zeta^3 & -\omega^2 & \omega^1 & 0 \end{pmatrix}. \tag{5.3}$$

We shall use also the notations $\omega = |\boldsymbol{\omega}|$ and $\zeta = |\boldsymbol{\zeta}|$. Notice that our matrix row and column indices run from 0 to 3. We find immediately $\det \mathbf{A} = [\boldsymbol{\omega}\cdot\boldsymbol{\zeta}]^2$. The characteristic polynomial,

$$\det[\lambda \mathbf{I} - \mathbf{A}] = \lambda^4 + [\omega^2 - \zeta^2]\,\lambda^2 - [\boldsymbol{\omega}\cdot\boldsymbol{\zeta}]^2, \tag{5.4}$$

shows that the two basic invariants are $\varphi_2 = \omega^2 - \zeta^2$ and $\varphi_4 = [\boldsymbol{\omega}\cdot\boldsymbol{\zeta}]^2$. For practical reasons we shall in what follows use

$$f_1 = \omega^2 - \zeta^2\,;\quad f_2 = \boldsymbol{\omega}\cdot\boldsymbol{\zeta}\,. \tag{5.5}$$

By the Cayley-Hamilton theorem, (5.4) says that $\mathbf{A}^4$ is not independent: $\mathbf{A}^4 = [\zeta^2 - \omega^2]\mathbf{A}^2 + [\boldsymbol{\omega}\cdot\boldsymbol{\zeta}]^2$. Consequently, all the higher powers of $\mathbf{A}$ are written in terms of $\mathbf{I}, \mathbf{A}, \mathbf{A}^2$ and $\mathbf{A}^3$. As $\operatorname{tr} \mathbf{A} = \operatorname{tr} \mathbf{A}^3 = 0$, it comes from (2.17) that $\det(\lambda \mathbf{I} - \mathbf{A}) = \lambda^4 - \dfrac{\lambda^2}{2!}\operatorname{tr}\mathbf{A}^2 + \dfrac{1}{4!}\{3(\operatorname{tr}\mathbf{A}^2)^2 - 3!\operatorname{tr}\mathbf{A}^4\}$ and from (2.11) that the last coefficient is related to $\det \mathbf{A}$: $\det \mathbf{A} = \dfrac{1}{4!}\{3(\operatorname{tr}\mathbf{A}^2)^2 - 3!\operatorname{tr}\mathbf{A}^4\}$. It will be convenient to introduce the two invariant expressions

$$U = \left[-f_1/2 + \left(f_1^2/4 + f_2^2\right)^{1/2}\right]^{1/2}; \tag{5.6}$$

$$V = \left[-f_1/2 - \left(f_1^2/4 + f_2^2\right)^{1/2}\right]^{1/2}. \tag{5.7}$$

They will turn up as the angles, real or imaginary, parametrizing the finite transformations. The values $V = 0$ and $U = \zeta$ will represent a pure boost; $U = 0$ and $V = i\,\omega$, a pure rotation. The four roots of the secular equation are found to be $\lambda_{1,2} = \pm U$ and $\lambda_{3,4} = \pm V$.

The finite transformation will be

$$\Lambda(\omega,\zeta) = e^{\mathbf{A}} = e^{\lambda_1}Z_1 + e^{\lambda_2}Z_2 + e^{\lambda_3}Z_3 + e^{\lambda_4}Z_4 =$$

$$(Z_1+Z_2)\cosh U + (Z_1-Z_2)\sinh U + (Z_3+Z_4)\cosh V + (Z_3-Z_4)\sinh V. \tag{5.8}$$

The projectors (3.5) come up as

$$Z_1 = \frac{(\mathbf{A}+U)(\mathbf{A}^2-V^2)}{2U(U^2-V^2)} \;;\; Z_2 = -\frac{(\mathbf{A}-U)(\mathbf{A}^2-V^2)}{2U(U^2-V^2)} \;;$$

$$Z_3 = \frac{(\mathbf{A}+V)(\mathbf{A}^2-U^2)}{2V(V^2-U^2)} \;;\; Z_4 = -\frac{(\mathbf{A}-V)(\mathbf{A}^2-U^2)}{2V(V^2-U^2)} \;.$$

It is then a matter of simple substitution to arrive at

$$\Lambda(\omega,\zeta) = \frac{1}{UV(U^2-V^2)}\left\{[\mathbf{A}^2 UV - \mathbf{I}UV^3]\cosh U + [\mathbf{A}^3 V - \mathbf{A}V^3]\sinh U\right.$$
$$\left. + [\mathbf{A}U^3 - \mathbf{A}^3 U]\sinh V + [\mathbf{I}U^3 V - \mathbf{A}^2 UV]\cosh V\right\}. \tag{5.9}$$

We have delayed the presentation of the necessary powers of A. Their computation is easier if we start by writing the entries in (5.3) as

$$\mathbf{A}_{ab} = -[\delta_{a0}\delta_{bi}\zeta_i + \delta_{ai}(\delta_{b0}\zeta_i + \delta_{bj}\varepsilon_{ijk}\omega_k)], \tag{5.10}$$

with a,b, ... = 0,1,2,3 and i,j,k = 1,2,3. This kind of notation is extremely convenient for explicit calculations. Furthermore, to get compact forms, it will be necessary to introduce a few objects: the vectors

$$\mathbf{Q} = \omega \times \zeta \;;$$
$$\mathbf{C} = (\omega^2 - \zeta^2)\,\omega + (\zeta\cdot\omega)\,\zeta = f_1\omega + f_2\zeta \;;$$
$$\mathbf{D} = (\omega^2 - \zeta^2)\zeta - (\zeta\cdot\omega)\,\omega = f_1\zeta - f_2\omega \;;$$

$$\mathbf{X} = V^2 \zeta + (\zeta \cdot \omega) \omega = V^2 \zeta + f_2 \omega \, ;$$
$$\mathbf{Y} = U^2 \zeta + (\zeta \cdot \omega) \omega = U^2 \zeta + f_2 \omega;$$
$$\mathbf{W} = V^2 \omega - f_2 \zeta \, ; \quad \mathbf{Z} = U^2 \omega - f_2 \zeta \, ; \tag{5.11}$$

and the bilinear
$$L_{ij} = \zeta_i \zeta_j + \omega_i \omega_j - \delta_{ij} \omega^2. \tag{5.12}$$

The powers of $\mathbf{A}$ will then have the entries
$$\mathbf{A}^2{}_{ab} = \delta_{a0}\delta_{b0}\zeta^2 + \delta_{a0}\delta_{bi}Q_i + \delta_{ai}\delta_{bj}L_{ij} - \delta_{ai}\delta_{b0}Q_i; \tag{5.13}$$

$$\mathbf{A}^3{}_{ab} = \delta_{a0}\delta_{bi}D_i + \delta_{ai}\delta_{b0}D_i + \delta_{ai}\delta_{bj}\varepsilon_{ijk}C_k. \tag{5.14}$$

We check immediately that tr $\mathbf{A}^3 = 0$ as expected, and find also $\text{tr}\mathbf{A}^2 = 2(\zeta^2 - \omega^2)$ and $\text{tr}\mathbf{A}^4 = 4[\omega \cdot \zeta]^2 + 2(\zeta^2 - \omega^2)^2$. The entries of (5.9) are then

$$\Lambda(\omega, \zeta)_{ab} = \left(e^{\mathbf{A}}\right)_{ab} = \frac{1}{UV(U^2-V^2)} \left\{ \mathbf{I}_{ab} [U^3 V \cosh V - UV^3 \cosh U] \right.$$
$$+ \mathbf{A}_{ab} [U^3 \sinh V - V^3 \sinh U] + \mathbf{A}^2{}_{ab} [UV \cosh U - UV \cosh V]$$
$$\left. + \mathbf{A}^3{}_{ab} [V \sinh U - U \sinh V] \right\}. \tag{5.15}$$

The result of substituting the powers is

$$\left(e^{\mathbf{A}}\right)_{ab} = \delta_{a0}\delta_{b0}\left\{\frac{\zeta^2 - V^2}{U^2-V^2} \cosh U + \frac{U^2 - \zeta^2}{U^2-V^2} \cosh V\right\}$$
$$+ \delta_{a0}\delta_{bi}\left\{\frac{X_i}{U^2-V^2} \sinh V/V - \frac{Y_i}{U^2-V^2} \sinh U/U - \frac{Q_i}{U^2-V^2}(\cosh V - \cosh U)\right\}$$
$$+ \delta_{ai}\delta_{b0}\left\{\frac{X_i}{U^2-V^2} \sinh V/V - \frac{Y_i}{U^2-V^2} \sinh U/U + \frac{Q_i}{U^2-V^2}(\cosh V - \cosh U)\right\}$$
$$+ \delta_{ai}\delta_{bj}\left\{\frac{L_{ij}-\delta_{ij}V^2}{U^2-V^2} \cosh U - \frac{L_{ij}-\delta_{ij}U^2}{U^2-V^2} \cosh V\right.$$
$$\left. + \varepsilon_{ijk}[\frac{W_k}{U^2-V^2} \sinh V/V - \frac{Z_k}{U^2-V^2} \sinh U/U]\right\} \tag{5.16}$$

The general Lorentz transformation will thus have the matrix form

$$\Lambda(\boldsymbol{\omega}, \boldsymbol{\zeta}) = e^{\mathbf{A}} = \frac{1}{U^2 - V^2} \times$$

$$\begin{pmatrix}
(\zeta^2 - V^2)\cosh U + (U^2 - \zeta^2)\cosh V & X_1 \frac{\sinh V}{V} - Y_1 \frac{\sinh U}{U} - Q_1(\cosh V - \cosh U) & X_2 \frac{\sinh V}{V} - Y_2 \frac{\sinh U}{U} - Q_2(\cosh V - \cosh U) & X_3 \frac{\sinh V}{V} - Y_3 \frac{\sinh U}{U} - Q_3(\cosh V - \cosh U) \\
X_1 \frac{\sinh V}{V} - Y_1 \frac{\sinh U}{U} + Q_1(\cosh V - \cosh U) & (\cosh U - \cosh V) L_{11} - V^2 \cosh U + U^2 \cosh V & (\cosh U - \cosh V) L_{12} + (\frac{\sinh U}{U} - \frac{\sinh V}{V})(\zeta \cdot \omega)\zeta_3 + (V\sinh V - U\sinh U)\omega_3 & (\cosh U - \cosh V) L_{13} - (\frac{\sinh U}{U} - \frac{\sinh V}{V})(\zeta \cdot \omega)\zeta_2 - (V\sinh V - U\sinh U)\omega_2 \\
X_2 \frac{\sinh V}{V} - Y_2 \frac{\sinh U}{U} + Q_2(\cosh V - \cosh U) & (\cosh U - \cosh V) L_{12} - (\frac{\sinh U}{U} - \frac{\sinh V}{V})(\zeta \cdot \omega)\zeta_3 - (V\sinh V - U\sinh U)\omega_3 & (\cosh U - \cosh V) L_{22} - V^2 \cosh U + U^2 \cosh V & (\cosh U - \cosh V) L_{23} + (\frac{\sinh U}{U} - \frac{\sinh V}{V})(\zeta \cdot \omega)\zeta_1 + (V\sinh V - U\sinh U)\omega_1 \\
X_3 \frac{\sinh V}{V} - Y_3 \frac{\sinh U}{U} + Q_3(\cosh V - \cosh U) & (\cosh U - \cosh V) L_{13} + (\frac{\sinh U}{U} - \frac{\sinh V}{V})(\zeta \cdot \omega)\zeta_2 + (V\sinh V - U\sinh U)\omega_2 & (\cosh U - \cosh V) L_{23} - (\frac{\sinh U}{U} - \frac{\sinh V}{V})(\zeta \cdot \omega)\zeta_1 - (V\sinh V - U\sinh U)\omega_1 & (\cosh U - \cosh V) L_{33} - V^2 \cosh U + U^2 \cosh V
\end{pmatrix}$$

(5.17)

We should be attentive to the limit cases, as the simple version of the method given here could fail if some eigenvalues become identical. It is enough to be careful. In the limit towards the identity, we find that for U and V small, the terms behave as they should: the diagonal terms as $\Lambda_{11} \to 1 + \zeta^2/2$; $\Lambda_{22} \to 1 + (L_{11} - \omega^2)/2$; the off-diagonal terms, as $\Lambda_{12} \to (Q_1/2 - \zeta_1)$; etc. We find just pure rotation for $\zeta = 0$, and the usual expression (Jackson, 1975) for a general pure boost when $\boldsymbol{\omega} = 0$.

## 6. POINCARÉ TRANSFORMATIONS

The inhomogeneous Lorentz transformation will have much in common with the homogeneous Lorentz case, but also some significant differences, stemming mainly from an extra (vanishing) root in the characteristic polynomial. In terms of the rotation $\boldsymbol{\omega}$, boost rapidities $\boldsymbol{\zeta}$, space translation parameters $\mathbf{a} = (a_1, a_2, a_3)$ and time translation parameter $a_0$, the basic algebra element will be

$$\mathbf{A} = \begin{pmatrix} 0 & -\zeta_1 & -\zeta_2 & -\zeta_3 & a_0 \\ -\zeta_1 & 0 & -\omega_3 & \omega_2 & a_1 \\ -\zeta_2 & \omega_3 & 0 & -\omega_1 & a_2 \\ -\zeta_3 & -\omega_2 & \omega_1 & 0 & a_3 \\ 0 & 0 & 0 & 0 & 0 \end{pmatrix}. \quad (6.1)$$

The characteristic polynomial will have two vanishing coefficients and can be conveniently written in the form

$$\lambda^5 + f_1 \lambda^3 - f_2^2 \lambda = 0, \tag{6.2}$$

which should be compared to (5.4). The two invariants are $f_1 = \omega^2 - \zeta^2$ and $f_2 = \omega \cdot \zeta$, the same as for the Lorentz case. Of the roots, four are also the same, just those of (5.6,7):

$$\lambda_1 = -\lambda_2 = U = \left[-f_1/2 + \left(f_1^2/4 + f_2^2\right)^{1/2}\right]^{1/2};$$
$$\lambda_3 = -\lambda_4 = V = \left[-f_1/2 - \left(f_1^2/4 + f_2^2\right)^{1/2}\right]^{1/2}. \tag{6.3}$$

There is, however an extra root,

$$\lambda_0 = 0. \tag{6.4}$$

In terms of the eigenprojectors, the general group element will be

$$g = e^0 Z_0 + e^U Z_1 + e^{-U} Z_2 + e^V Z_3 + e^{-V} Z_4 =$$
$$= Z_0 + \sinh U(Z_1 - Z_2) + \cosh U(Z_1 + Z_2) + \sinh V(Z_3 - Z_4) + \cosh V(Z_3 + Z_4). \tag{6.5}$$

Taking into account the relationships between the roots and choosing convenient denominators, the projectors become

$$Z_0 = I + \frac{\mathbf{A}^4 - \mathbf{A}^2(U^2 + V^2)}{U^2 V^2} \; ; \quad Z_1 = \frac{\mathbf{A}^4 V^2 + \mathbf{A}^3 U V^2 - \mathbf{A}^2 V^4 - \mathbf{A} U V^4}{2U^2 V^2 (U^2 - V^2)} \; ;$$

$$Z_2 = \frac{\mathbf{A}^4 V^2 - \mathbf{A}^3 U V^2 - \mathbf{A}^2 V^4 + \mathbf{A} U V^4}{2U^2 V^2 (U^2 - V^2)} \; ; Z_3 = \frac{-\mathbf{A}^4 U^2 + \mathbf{A}^2 U^4 - \mathbf{A}^3 U^2 V + \mathbf{A} U^4 V}{2U^2 V^2 (U^2 - V^2)} \; ;$$

$$Z_4 = \frac{-\mathbf{A}^4 U^2 + \mathbf{A}^2 U^4 + \mathbf{A}^3 U^2 V - \mathbf{A} U^4 V}{2U^2 V^2 (U^2 - V^2)} \; .$$

In terms of powers of $\mathbf{A}$, the general group element will be

$$g = I + \frac{\mathbf{A}^4 - \mathbf{A}^2(U^2 + V^2)}{U^2 V^2} +$$

$$\frac{1}{U^2-V^2} \left\{ \frac{A^3-AV^2}{U} \sinh U + \frac{A^4-A^2V^2}{U^2} \cosh U + \frac{AU^2-A^3}{V} \sinh V + \frac{A^2U^2-A^4}{V^2} \cosh V \right\}$$
(6.6)

with entries

$$g_{ab} = \mathbf{I}_{ab} + \frac{1}{(U^2-V^2)U^2V^2} \left\{ [U^4 V \sinh V - UV^4 \sinh U] \mathbf{A}_{ab} + [U^4 \cosh V - V^4 \cosh U - (U^4-V^4)] \mathbf{A}^2_{ab} + [UV^2 \sinh U - U^2 V \sinh V] \mathbf{A}^3_{ab} + [(U^2-V^2) + V^2 \cosh U - U^2 \cosh V] \mathbf{A}^4_{ab} \right\}.$$
(6.7)

We need now the detailed expressions of the powers of A. The entries in (6.1) are

$$A_{ab} = -\delta_{a0}\delta_{bi}\zeta_i - \delta_{ai}(\delta_{b0}\zeta_i + \delta_{bj}\varepsilon_{ijk}\omega_k) + \delta_{a0}\delta_{b4}a_0 + \delta_{ai}\delta_{b4}a_i,$$
(6.8)

with now a,b, ... = 0,1,2,3,4 and i,j,k = 1,2,3. Using (5.11-12), plus the vectors

$$\mathbf{P} = \boldsymbol{\omega} \times \mathbf{a} - a_0 \boldsymbol{\zeta};$$
$$\mathbf{K} = -\omega^2 \mathbf{a} + (\mathbf{a} \cdot \boldsymbol{\omega})\boldsymbol{\omega} + (\mathbf{a} \cdot \boldsymbol{\zeta})\boldsymbol{\zeta} - a_0 \mathbf{Q};$$
$$\mathbf{M} = \mathbf{a} \times \mathbf{C} + a_0 \mathbf{D},$$
(6.9)

the powers of A will have the matrix elements

$$A^2_{ab} = \delta_{a0}\delta_{b0}\zeta^2 + \delta_{a0}\delta_{bi}Q_i - \delta_{a0}\delta_{b4}\boldsymbol{\zeta}\cdot\mathbf{a} + \delta_{ai}\delta_{bj}L_{ij} + \delta_{ai}\delta_{b4}P_i - \delta_{ai}\delta_{b0}Q_i;$$

$$A^3_{ab} = \delta_{a0}\delta_{bi}D_i - \delta_{a0}\delta_{b4}\boldsymbol{\zeta}\cdot\mathbf{P} + \delta_{ai}\delta_{b0}D_i + \delta_{ai}\delta_{b4}K_i + \delta_{ai}\delta_{bj}\varepsilon_{ijk}C_k;$$

$$A^4_{ab} = -\delta_{a0}\delta_{b0}\boldsymbol{\zeta}\cdot\mathbf{D} - \delta_{a0}\delta_{bi}f_1 Q_i + \delta_{a0}\delta_{b4}\mathbf{a}\cdot\mathbf{D} + \delta_{ai}\delta_{b4}M_i$$
$$+ \delta_{ai}\delta_{b0}f_1 Q_i + \delta_{ai}\delta_{bj}[\delta_{ij}f_2^2 - f_1 L_{ij}].$$
(6.10)

In matrix form, they are

$$A^2 = \begin{pmatrix} \zeta^2 & Q_1 & Q_2 & Q_3 & -\mathbf{a}\cdot\boldsymbol{\zeta} \\ -Q_1 & L_{11} & L_{12} & L_{13} & P_1 \\ -Q_2 & L_{12} & L_{22} & L_{23} & P_2 \\ -Q_3 & L_{13} & L_{23} & L_{33} & P_3 \\ 0 & 0 & 0 & 0 & 0 \end{pmatrix} ; A^3 = \begin{pmatrix} 0 & D_1 & D_2 & D_3 & -\boldsymbol{\zeta}\cdot\mathbf{P} \\ D_1 & 0 & C_3 & -C_2 & K_1 \\ D_2 & -C_3 & 0 & C_1 & K_2 \\ D_3 & C_2 & -C_1 & 0 & K_3 \\ 0 & 0 & 0 & 0 & 0 \end{pmatrix} ;$$

$$A^4 = \begin{pmatrix} f_2^2 - f_1\zeta^2 & -f_1 Q_1 & -f_1 Q_2 & -f_1 Q_3 & \mathbf{a}\cdot\mathbf{D} \\ f_1 Q_1 & f_2^2 - f_1 L_{11} & -f_1 L_{12} & -f_1 L_{13} & M_1 \\ f_1 Q_2 & -f_1 L_{12} & f_2^2 - f_1 L_{22} & -f_1 L_{23} & M_2 \\ f_1 Q_3 & -f_1 L_{13} & -f_1 L_{23} & f_2^2 - f_1 L_{33} & M_3 \\ 0 & 0 & 0 & 0 & 0 \end{pmatrix}. \quad (6.11)$$

It is easily checked that, as expected from (2.18), $\mathrm{tr} A^2 = -2f_1$ and $\mathrm{tr} A^4 = 2(U^4+V^4) = 2f_1^2 + 4f_2^2$.

Computation of the generic Poincaré group matrix element now can be made from (6.7). Direct calculation yields the rather awkward general form. We can, however, use identities such as $U^2+V^2 = -f_1$ and $U^2 V^2 = -f_2^2$, as well as the definitions of the various vectors, to get simpler expressions. We find identities like $(U^2+V^2)\zeta^2 + \zeta\cdot\mathbf{D} - U^2 V^2 = 0$; $V^2\zeta + \mathbf{D} = -\mathbf{Y}$; $U^2\zeta + \mathbf{D} = -\mathbf{X}$; $V^2\boldsymbol{\omega} - \mathbf{C} = \mathbf{W} - f_1\boldsymbol{\omega}$; $\mathbf{C} - U^2\boldsymbol{\omega} = f_1\boldsymbol{\omega} - \mathbf{Z}$; etc. There are, so, many different though equivalent expressions for the entries. One of the simplest is the following:

$$g_{ab} = \delta_{a4}\delta_{b4}$$
$$+ \frac{1}{(U^2-V^2)U^2V^2}\Big\{\delta_{a0}\delta_{b0}\big[(\zeta^2-V^2)U^2V^2\cosh U - (\zeta^2-U^2)U^2V^2\cosh V\big]$$
$$+ \delta_{a0}\delta_{bi}\big[X_i U^2 V \sinh V - Y_i U V^2 \sinh U + U^2 V^2 Q_i(\cosh U - \cosh V)\big]$$
$$+ \delta_{ai}\delta_{b0}\big[X_i U^2 V \sinh - Y_i U V^2 \sinh U - U^2 V^2 Q_i(\cosh U - \cosh V)\big]$$
$$+ \delta_{ai}\delta_{b4}\big[(K_i - V^2 a_i)UV^2 \sinh U + (U^2 a_i - K_i)U^2 V \sinh V + (U^2 P_i - M_i)U^2 \cosh V - (V^2 P_i - M_i)V^2 \cosh U + (U^2-V^2)(M_i + f_1 P_i)\big]$$
$$+ \delta_{a0}\delta_{b4}\big[(U^2 a_0 + \zeta\cdot\mathbf{P})U^2 V \sinh V - (V^2 a_0 + \zeta\cdot\mathbf{P})U V^2 \sinh U + \mathbf{a}\cdot(\mathbf{X}U^2 \cosh V - \mathbf{Y}V^2 \cosh U - f_2(U^2-V^2)\boldsymbol{\omega})\big]$$
$$+ \delta_{ai}\delta_{bj}\big[(U^2 L_{ij} + \delta_{ij}f_2^2)V^2 \cosh U - (V^2 L_{ij} + \delta_{ij}f_2^2)U^2 \cosh V - \varepsilon_{ijk} Z_k U V^2 \sinh U + \varepsilon_{ijk} W_k U^2 V \sinh V\big]\Big\}. \quad (6.12)$$

The entries in common are exactly those in (5.15). And here we arrive at a limitation of the method. Proceeding as described, we do obtain the generic group element as the exponential of the generic algebra member, but the result is very complicated. The (space and time) translation parameters appear mixed in a very cumbersome way. A detailed examination of the Galilei case shows the reason for that. There, direct exponentiation is feasible, and the study of the order-by-order effects of one transformation on the other shows how, at each order, the parameters of one transformation is modified by the other transformations. The result is not interesting, as the usual parametrizations are much simpler. Recall that there are two possible simple parametrizations for semidirect products like the Euclidean and the Poincaré groups. This is due to the fact that a general transformation is the product of a rotation (in 3-space or in spacetime, respectively) by a translation (idem), and that this product can be chosen to be done in the inverse order, a translation by a rotation. An inhomogeneous Lorentz transformation $P = (L,a)$ can be given either by $x' = L(x+a)$ or by $x' = Lx+a$. In the first case translations are performed first and the homogeneous Lorentz transformations will act also on them. The latter is the more usual parametrization,

$$g(\boldsymbol{\omega},\boldsymbol{\zeta},\mathbf{a}) = \begin{pmatrix} \Lambda_{00} & \Lambda_{01} & \Lambda_{02} & \Lambda_{03} & a_0 \\ \Lambda_{10} & \Lambda_{11} & \Lambda_{12} & \Lambda_{13} & a_1 \\ \Lambda_{20} & \Lambda_{21} & \Lambda_{22} & \Lambda_{23} & a_2 \\ \Lambda_{30} & \Lambda_{31} & \Lambda_{32} & \Lambda_{33} & a_3 \\ 0 & 0 & 0 & 0 & 1 \end{pmatrix}. \tag{6.13}$$

where $\Lambda(\boldsymbol{\omega}, \boldsymbol{\zeta})$ indicates the matrix (5.16). We can obtain it from the exponentiated algebra member, but a fairly involved redefinition of the parameters is necessary. The idea will be shown in case of the Galilei group, which is, in this particular aspect, simpler.

## 7. THE GALILEI GROUP

The application of the method of matrix functions is, for the Galilei group, far more complicated: the matrix representing the algebra member has coincident eigenvalues. With the rotation angles collected as above in the vector $\omega = (\omega_1,\omega_2,\omega_3)$, velocity $\mathbf{v}$, space translation parameters $\mathbf{a} = (a_1,a_2,a_3)$ and time translation parameter $a_0$, the generic member of the Galilei algebra is

$$A = \begin{pmatrix} 0 & 0 & 0 & 0 & a_0 \\ -v_1 & 0 & -\omega_3 & \omega_2 & a_1 \\ -v_2 & \omega_3 & 0 & -\omega_1 & a_2 \\ -v_3 & -\omega_2 & \omega_1 & 0 & a_3 \\ 0 & 0 & 0 & 0 & 0 \end{pmatrix} \tag{7.1}$$

Notice det A = 0, tr A = 0, from which we expect $\varphi_5 = 0$ and $\varphi_1 = 0$. The characteristic polynomial will be $\det[\lambda I - A] = \lambda^3(\lambda^2 + \omega^2)$, so that the whole list of expected invariants is $\varphi_0 = 1$; $\varphi_1 = 0$; $\varphi_2 = \omega^2$; $\varphi_3 = 0$; $\varphi_4 = 0$; $\varphi_5 = 0$. The secular equation

$$\lambda^3(\lambda^2 + \omega^2) = 0 \tag{7.2}$$

has the roots $\lambda_1 = \lambda_2 = \lambda_3 = 0$; $\lambda_4 = +i\omega$ and $\lambda_5 = -i\omega$. There is only one invariant, the same $\omega^2$ of the rotation subgroup. The Cayley-Hamilton theorem gives here very simple expressions for the higher powers, $A^5 = -\omega^2 A^3$, etc, so that it is actually very easy to get the generic group element $G(\omega,\mathbf{v},a) = \exp[A]$ by direct exponentiation. We shall, however, use this case to illustrate the general method, taking into account the triple eigenvalue degeneracy. The expression (3.10) takes the form

$$\begin{vmatrix} F(A) & F(\lambda_1) & F'(\lambda_1) & F''(\lambda_1) & F(\lambda_4) & F(\lambda_5) \\ I & 1 & 0 & 0 & 1 & 1 \\ A & 0 & 1 & 0 & \lambda_4 & \lambda_5 \\ A^2 & 0 & 0 & 2 & \lambda_4^2 & \lambda_5^2 \\ A^3 & 0 & 0 & 0 & \lambda_4^3 & \lambda_5^3 \\ A^4 & 0 & 0 & 0 & \lambda_4^4 & \lambda_5^4 \end{vmatrix} = 0. \tag{7.3}$$

Putting on the values of interest for us, it becomes

$$\begin{vmatrix} e^A & 1 & 1 & 1 & e^{i\omega} & e^{-i\omega} \\ I & 1 & 0 & 0 & 1 & 1 \\ A & 0 & 1 & 0 & i\omega & -i\omega \\ A^2 & 0 & 0 & 2 & -\omega^2 & -\omega^2 \\ A^3 & 0 & 0 & 0 & -i\omega^3 & i\omega^3 \\ A^4 & 0 & 0 & 0 & \omega^4 & \omega^4 \end{vmatrix} = 0. \tag{7.4}$$

Taking the coefficients in the expansion along the first row, we find the operators taking the places of would-be projectors:

$$Z_{10} = I - \frac{A^4}{\omega^4} \; ; Z_{11} = A(I + \frac{A^2}{\omega^2}) ; Z_{12} = \frac{A^2}{2}(I + \frac{A^2}{\omega^2}) ;$$

$$Z_4 = \frac{A^2}{2\omega^2}(\frac{A}{\omega} + i) \; ; Z_5 = \frac{A^2}{2\omega^2}(\frac{A}{\omega} - i) .$$

$Z_4$ and $Z_5$ are real projectors, idempotents and annihilating each other. But $Z_{10}$, $Z_{11}$ and $Z_{12}$ are not. They do satisfy the relations $Z_4 Z_{10} = 0$; $Z_5 Z_{10} = 0$; $Z_{10}^2 = Z_{10}$; and $Z_{10} + Z_4 + Z_5 = I$. But they fail to satisfy the other necessary relations to be projectors. Thus, for example, $Z_{10}Z_{11} = Z_{11}$; $Z_{10}Z_{12} = Z_{12}$; $Z_{11}Z_{12} = 0$; $Z_{12}^2 = 0$; etc.

Expanding the formal determinant along the first column, we obtain

$$e^A = I + A + \frac{A^2}{2} + (\frac{A}{\omega})^4 (\frac{\omega^2}{2} + \cos\omega - 1) + (\frac{A}{\omega})^3 (\omega - \sin\omega). \quad (7.5)$$

In order to display the powers of A, it will be convenient to define, besides the unit $\mathbf{u} = \omega/|\omega|$, the 3-vectors $\mathbf{q} = \omega \times \mathbf{v}$; $\mathbf{p} = \omega \times \mathbf{a}$. And we shall also use, for the vectors parallel and transversal to $\mathbf{u}$, the notations $\mathbf{a}^{\|} = (\mathbf{a}\cdot\mathbf{u})\mathbf{u}$, $\mathbf{v}^{\|} = (\mathbf{v}\cdot\mathbf{u})\mathbf{u}$ and $\mathbf{a}^{\perp} = \mathbf{a} - (\mathbf{a}\cdot\mathbf{u})\mathbf{u}$, $\mathbf{v}^{\perp} = \mathbf{v} - (\mathbf{v}\cdot\mathbf{u})\mathbf{u}$ respectively. Then, always labeling "0" the first row and column we have, for a,b = 0, 1,2,3,4:

$$A^2 = \begin{pmatrix} 0 & 0 & 0 & 0 & 0 \\ -q_1 & -\omega_2^2-\omega_3^2 & \omega_1\omega_2 & \omega_1\omega_3 & p_1-a_o v_1 \\ -q_2 & \omega_1\omega_2 & -\omega_1^2-\omega_3^2 & \omega_2\omega_3 & p_2-a_o v_2 \\ -q_3 & \omega_1\omega_3 & \omega_2\omega_3 & -\omega_1^2-\omega_2^2 & p_3-a_o v_3 \\ 0 & 0 & 0 & 0 & 0 \end{pmatrix}. \quad (7.6)$$

We see that tr $A^2 = -2\omega^2$, coherent with $\varphi_2 = \omega^2$. We find next

$$A^3 = \omega^2 \begin{pmatrix} 0 & 0 & 0 & 0 & 0 \\ v^{\perp}_1 & 0 & \omega_3 & -\omega_2 & -a^{\perp}_1 - (a_o/\omega^2)q_1 \\ v^{\perp}_2 & -\omega_3 & 0 & \omega_1 & -a^{\perp}_2 - (a_o/\omega^2)q_2 \\ v^{\perp}_3 & \omega_2 & -\omega_1 & 0 & -a^{\perp}_3 - (a_o/\omega^2)q_3 \\ 0 & 0 & 0 & 0 & 0 \end{pmatrix}; \quad (7.7)$$

$$A^4 = \omega^2 \begin{pmatrix} 0 & 0 & 0 & 0 & 0 \\ q_1 & \omega^2-\omega_1^2 & -\omega_1\omega_2 & -\omega_1\omega_3 & a_o v^\perp_1 - p_1 \\ q_2 & -\omega_1\omega_2 & \omega^2-\omega_2^2 & -\omega_2\omega_3 & a_o v^\perp_2 - p_2 \\ q_3 & -\omega_1\omega_3 & -\omega_2\omega_3 & \omega^2-\omega_3^2 & a_o v^\perp_3 - p_3 \\ 0 & 0 & 0 & 0 & 0 \end{pmatrix}. \quad (7.8)$$

We find the expected results tr $A^3 = 0$ and tr $A^4 = 2\,\omega^4$, consistent with $\varphi_4 = 0$. It is convenient, both to perform the calculations and to exhibit the resulting large matrices, to display the matrix elements in the forms

$$A_{ab} = \delta_{a0}\delta_{b4}a_0 + \delta_{ai}[\delta_{b4}a_i - \delta_{b0}v_i - \delta_{bj}\varepsilon_{ijk}\omega_k];$$

$$A^2{}_{ab} = -\delta_{ai}[\delta_{b4}(a_0 v_i - p_i) + \delta_{b0}q_i + \delta_{bj}(\delta_{ij}\omega^2 - \omega_i\omega_j)];$$

$$A^3{}_{ab} = \delta_{ai}\omega^2 \left\{ -\delta_{b4}\left[\frac{a_0}{\omega^2}\,q_i + a^\perp_i\right] + \delta_{b0}v^\perp_i + \delta_{bj}\varepsilon_{ijk}\omega_k \right\};$$

$$A^4{}_{ab} = \delta_{ai}\omega^4 \left\{ \delta_{b4}\left[\frac{a_0 v^\perp_i + (\mathbf{a}\times\boldsymbol\omega)_i}{\omega^2}\right] + \delta_{b0}\left(\frac{\mathbf{u}\times\mathbf{v}}{\omega}\right)_i + \delta_{bj}(\delta_{ij} - u_i u_j) \right\}.$$

The Galilei group element comes out then as

$$[e^A]_{ab} = \delta_{a0}\delta_{b0} + \delta_{a4}\delta_{b4} + \delta_{a0}\delta_{b4}a_0 + \delta_{ai}\delta_{bj}R_{ij}$$

$$+ \delta_{ai}\left\{ \delta_{b4}\left[ a^\parallel_i - \tfrac{1}{2}a_0 v^\parallel_i + a^\perp_i \frac{\sin\omega}{\omega} + (\mathbf{a}\times\mathbf{u})_i \frac{\cos\omega - 1}{\omega} \right.\right.$$

$$\left.\left. + \frac{a_0}{\omega}(\mathbf{v}\times\mathbf{u})_i \left[1 - \frac{\sin\omega}{\omega}\right] + \frac{a_0 v^\perp_i}{\omega^2}(\cos\omega - 1) \right]\right.$$

$$\left. - \delta_{b0}\left[ v^\parallel_i - (\mathbf{v}\times\mathbf{u})_i \frac{1-\cos\omega}{\omega} + v^\perp_i \frac{\sin\omega}{\omega} \right] \right\}. \quad (7.9)$$

where $R_{ij}$ are the rotation group entries $R_{ij} = \delta_{ij} - \varepsilon_{ijk}u_k \sin\omega + (u_i u_j - \delta_{ij})(1 - \cos\omega)$. Both the translation sector and the Galilean boosts are very complicated. The relation with the usual parametrization is seen if we redefine new parameters as $a'_i = (e^A)_{i4}$ and $v'_i = -(e^A)_{i4}$. Only then will the Galilei group element take the usual form

$$G(\omega,\mathbf{v},\mathbf{a}) = \begin{pmatrix} 1 & 0 & 0 & 0 & a_0 \\ -v'_1 & R_{11} & R_{12} & R_{13} & a'_1 \\ -v'_2 & R_{21} & R_{22} & R_{23} & a'_2 \\ -v'_3 & R_{31} & R_{32} & R_{33} & a'_3 \\ 0 & 0 & 0 & 0 & 1 \end{pmatrix}, \qquad (7.10)$$

The effect of exponentiation is analogous to that we have found in the Poincaré case: the translation and velocity parameters we started with get all mixed up. Another way to obtain the expressions for the Galilei group is by the Inönü-Wigner contraction (Gilmore, 1974)) of the Poincaré result. This would need a previous preparation of the matrices, with some extra factors of c and 1/c. It is simpler to start from another parametrization of the Poincaré algebra element: instead of (6.1), we take A changed by a similarity transformation, S A S$^{-1}$, with S = diag(1,c,c,c,1):

$$A = \begin{pmatrix} 0 & -\zeta_1/c & -\zeta_2/c & -\zeta_3/c & a_0 \\ -c\zeta_1 & 0 & -\omega_3 & \omega_2 & ca_1 \\ -c\zeta_2 & \omega_3 & 0 & -\omega_1 & ca_2 \\ -c\zeta_3 & -\omega_2 & \omega_1 & 0 & ca_3 \\ 0 & 0 & 0 & 0 & 0 \end{pmatrix}. \qquad (7.11)$$

This corresponds to dividing the entries (0k) of A by c and multiplying the entries (k0) by c. All the powers of A and the exponential itself will acquire the same factors in the corresponding entries. The result for the Galilei group comes by taking the limit c → ∞ though, as usual in the contraction procedure, the translation parameters must absorb a factor "c" in the limit.

## 8. FINAL COMMENTS

To arrive at the exponentials we could use directly equations (3.3) and (3.5) and ignore the treatment involving the characteristic polynomial and symmetric functions, leading to the closed expressions for the projectors. That discussion provides, however, a deeper insight into the whole subject, in particular showing how the generalized boost and rotation angles are invariant. The method requires the knowledge of the roots of the characteristic polynomial. The approach also throws a bridge towards the tantalizing recent results on the relationship between invariant polynomials and von Neumann algebras (Jones,1991). We have taken as granted that functions of matrices, as long as they can be defined, are completely determined by their spectra. This is justified in a much more general context. Matrix algebras are very particular kinds of von Neumann algebras and it is a very strong result of the theory of Banach algebras (Kirillov, 1974; Bratelli and Robinson,1979) that functions on such spaces, as long as defined, are indeed fixed by the spectra.

**Acknowledgments**

The authors are grateful to J.G. Pereira and L.C.B. Crispino for many discussions. They also would like to thank CNPq, Brasilia, for financial support.

**Appendix**

**SYMMETRIC FUNCTIONS**

A symmetric function (Littlewood, 1950; Comtet, 1974; MacDonald, 1979) in N variables $x_1, x_2, x_3, \ldots, x_N$ is a polynomial $F(x_1, x_2, x_3, \ldots, x_N)$ which is invariant under any permutation of the $x_k$'s. Their main properties are listed in appendix A. The variables may be called "letters" and we shall indicate them collectively by **x**. This set $\mathbf{x} = \{x_1, x_2, x_3, \ldots, x_N\}$ is called, naturally enough, the "alphabet" and a monomial is a "word". A symmetric function in N variables $F(x_1, x_2, x_3, \ldots, x_N)$ i $F(\mathbf{x})$ can be written as sum of words,

$$F(\mathbf{x}) = \Sigma_{\{n_1,n_2,n_3,\ldots\},\{i_1,i_2,i_3,\ldots,i_N\}} x_{i_1}^{n_1} x_{i_2}^{n_2} x_{i_3}^{n_3} \ldots x_{i_r}^{n_r}. \qquad (A.1)$$

Only two kinds of such functions will interest us:

(i) the *j-th elementary symmetric function* $\sigma_j$, the sum of all distinct words with j distinct letters:

$$\sigma_j[\mathbf{x}] = \Sigma^{(N)} x_1 x_2 \ldots x_j. \qquad (A.2)$$

As examples,

$\sigma_1[x] = x_1 + x_2 + x_3 + \ldots + x_N$;

$$\sigma_2[\mathbf{x}] = x_1 x_2 + x_1 x_3 + \ldots + x_1 x_N + \ldots + x_2 x_3 + x_2 x_4 + \ldots + \ldots + x_{N-1} x_N;$$

$$\ldots$$

$$\sigma_N[\mathbf{x}] = x_1 x_2 x_3 \ldots x_{N-1} x_N .$$

When the $x_j$'s are the eigenvalues of an N×N matrix A, clearly $\sigma_N[\mathbf{x}] = \det A$.

(ii) the *k-th power-sum symmetric function*

$$s_k[\mathbf{x}] = \sum_{j=1}^{N} x_j^k = (x_1)^k + (x_2)^k + (x_3)^k + \ldots + (x_N)^k . \tag{A.3}$$

When the $x_j$'s are the eigenvalues of A, $s_k[\mathbf{x}]$ is clearly the trace of $A^k$. These symmetric functions have the generating functions

$$P([\mathbf{x}], t) = \sum_{j=0}^{N} \sigma_j[\mathbf{x}] t^j = \prod_{j=1}^{N} (1 + x_j t) \tag{A.4}$$

and

$$\psi([\mathbf{x}], t) = \sum_{j=1}^{N} s_j[\mathbf{x}] t^{j-1} = \prod_{j=1}^{N} x_j (1 - x_j t)^{-1} , \tag{A.5}$$

with $\sigma_0 = 1$ by convention. The so called fundamental theorem of the symmetric functions says that the $\sigma_k$'s are algebraically independent. This means that, up to a constant, and for $N \to \infty$, a general formal series f(t) can be seen as the generating function $P([\mathbf{x}], t)$ with a convenient choice of the alphabet $\mathbf{x}$. There are two relationships (offsprings of the obvious relation $\psi([\mathbf{x}], t) = -\frac{d}{dt} \ln P([\mathbf{x}], -t)$), inverse to each other, between the functions above:

$$\sigma_j[\mathbf{x}] = \frac{1}{j!} \sum_{m=0}^{j} B_{jm} \{ (-)^{k-1} (k-1)! \, s_k[\mathbf{x}] \} ; \tag{A.6}$$

$$s_n[\mathbf{x}] = \frac{(-)^{n-1}}{(n-1)!} \sum_{m=0}^{n} (-)^{m-1} (m-1)! \, B_{nm} \{ k! \sigma_k[\mathbf{x}] \} . \tag{A.7}$$

Comparing (A.6) with (2.10), we see that the invariants $\varphi_j$ are (up to eventual signs) nothing more than symmetric functions of the eigenvalue-alphabet $\lambda =$ Sp A = $\{\lambda_1, \lambda_2, \lambda_3, \ldots \lambda_N\}$ of the generic Lie algebra member:

$$\varphi_j[A] = (-)^j \sigma_j[\lambda] \, . \tag{A.8}$$

With (A.7), we have so to say completed a circle: the $s_n[\mathbf{x}]$'s are just the original powers of traces. It is interesting to consider simultaneously two related alphabets. Take $\mathbf{x} = \{x_1, x_2, x_3, \ldots, x_N\}$ as above and call its "reciprocal" the alphabet $\mathbf{x}^* = \{x^*_1, x^*_2, x^*_3, \ldots, x_N^*\}$, where each $x^*_j = -1/x_j$. Notice that $x^{**}_k = x_k$. The $\sigma_k$'s of these alphabets are related by

$$(-)^j \sigma_{N-j}[\mathbf{x}] = \sigma_N[\mathbf{x}] \sigma_j[\mathbf{x}^*]. \tag{A.9}$$

The various forms of a general polynomial of roots $x_1, x_2, x_3, \ldots, x_N$ can be summed up in

$$P(t) = \sum_{j=0}^{N} b_{N-j} t^j = (-)^N \sigma_N[\mathbf{x}] \sum_{j=0}^{N} \sigma_j[\mathbf{x}^*] \, t^j = \sum_{j=0}^{N} (-)^j \sigma_j[\mathbf{x}] \, t^{N-j} =$$

$$\sum_{j=0}^{N} (-)^{N-j} \sigma_{N-j}[\mathbf{x}] \, t^j = (-)^N \sigma_N[\mathbf{x}] \prod_{j=1}^{N} (1 + x^*_j t) = (-)^N \sigma_N[\mathbf{x}] \prod_{j=1}^{N} (1 - t/x_j) =$$

$$= \prod_{j=1}^{N} (t - x_j) = P(0) \prod_{j=1}^{N} (1 - t/x_j) \, , \tag{A.10}$$

the latter being Weierstrass' expression for a polynomial in terms of its zeros. In the main text we make a particular use of the relation

$$\prod_{j=1}^{N} (x_j - t) = \sigma_N[\mathbf{x}] \sum_{j=0}^{N} \sigma_j[\mathbf{x}^*] t^j \, . \tag{A.11}$$

In the above expressions, the highest-order coefficient is $b_0 = 1$ and the independent term is $b_N = P(0) = (-)^N \sigma_N[\mathbf{x}]$. The general Vieta relation between the coefficients and the roots of a polynomial comes out immediately:

$$b_r = (-)^r \sigma_r[\mathbf{x}] = \frac{(-)^r}{r!} \sum_{m=0}^{r} B_{rm}\{(-)^{k-1}(k-1)! s_k[\mathbf{x}]\}. \tag{A.12}$$

We are of course concerned here with the classical case of commuting alphabets. Recent research on quantum groups and related symmetric functions of non-commutative variables (Gelfand 1994, 1995) have called attention to the interest of having closed forms like those above, even for the classical case. Let now $\sigma_{ji}[\mathbf{x}]$ be the sum of all j-products of the alphabet x, but excluding the letter $x_i$. For example, $\sigma_{Ni}[\mathbf{x}] = \prod_{k \neq i}^{N} x_k$. We put by convention $\sigma_{0i} = 1$ and find that

$$\sigma_{ki}[\mathbf{x}] = \sum_{p=0}^{k} (-)^p x_i^{\,p} \sigma_{k-p}[\mathbf{x}] = \sum_{j=0}^{k} (-x_i)^{k-j} \sigma_j[\mathbf{x}] \ . \tag{A.13}$$

In the absence of the i-th letter, (A.11) will become

$$\prod_{j=1; j \neq i}^{N} (x_j - t) = \sigma_{Ni}[\mathbf{x}] \sum_{k=0}^{N} \sigma_{ki}[\mathbf{x}^*] t^k = \sigma_{Ni}[\mathbf{x}] \sum_{k=0}^{N} \sum_{j=0}^{k} (-x_i)^{k-j} \sigma_j[\mathbf{x}^*] t^k.$$

Using (A.9),

$$\prod_{j=1; j \neq i}^{N} (x_j - t) = \frac{\sigma_{Ni}[\mathbf{x}]}{\sigma_N[\mathbf{x}]} \sum_{k=0}^{N} \sum_{j=0}^{k} (-x_i)^k \sigma_{N-j}[\mathbf{x}] \, t^k. \tag{A.14}$$

The projector $Z_i$ for a matrix A with eigenvalues forming the alphabet $\lambda = \{\lambda_1, \lambda_2, \ldots, \lambda_N\}$ can then be written as the ratio of two such polynomials, with variable $t = A$ and $t = \lambda_i$:

$$Z_i[A] = \prod_{k=1; k \neq i}^{N} \frac{\lambda_k - A}{\lambda_k - \lambda_i} = \frac{\sum_{k=0}^{N} \sigma_{ki}[\lambda^*] A^k}{\sum_{n=0}^{N} \sigma_{ni}[\lambda^*] \lambda_i^n} = \frac{\sum_{k=0}^{N} \sum_{j=0}^{k} \lambda_i^{j-k} \sigma_j[\lambda^*] A^k}{\sum_{n=0}^{N} \sum_{j=0}^{n} (\lambda_i)^j \sigma_j[\lambda^*]}$$

$$= \frac{\sum_{k=0}^{N} \sum_{j=0}^{k} \lambda_i^{j-k} (-)^j \sigma_{N-j}[\lambda] A^k}{\sum_{n=0}^{N} \sum_{j=0}^{n} (-\lambda_i)^j \sigma_{N-j}[\lambda]} = \sum_{k=0}^{N} \frac{\sum_{j=0}^{k} \lambda_i^{j-k} \varphi_{N-j}[A]}{\sum_{n=0}^{N} \sum_{j=0}^{n} \lambda_i^j \varphi_{N-j}[A]} A^k \ . \tag{A.15}$$

# REFERENCES

Aldrovandi R., and Monte Lima, I. (1980). J.Phys. **A13** 3685.

Aldrovandi R., and Monte Lima, I. (1983). Astrophys.&Space Sci. **90** 179.

Aldrovandi R., and Pereira, J.G. (1995). *Geometrical Physics*, World Scientific, Singapore.

Aldrovandi R., and E. Stedile, E. (1984). Int.J.Theor.Phys. **23** 301.

Bratelli, O., and Robinson, D.W. (1979). *Operator Algebras and Quantum Statistical Mechanics* 1, Springer, New York.

Comtet, L. (1974). *Advanced Combinatorics*, Reidel, Dordrecht.

Eguchi, T., Gilkey, P.B., and Hanson, A.J. (1980). Phys.Rep. **66** 213

Faddeev, D.K., and Faddeeva, V.N. (1963). *Computational Methods of Linear Algebra*, W.H. Freeman and Co., San Francisco.

Gantmacher, F.R. (1990). *The Theory of Matrices*, Chelsea Pub.Co., New York.

Gelfand, I.M., Krob, D., Lascoux, A., Leclerc, B., Retakh, V.S., and Thibon, J.-Y. (1994). *Noncommutative Symmetric Functions*, HEP-TH/9407124, preprint LITP 94.39.

Gelfand, I.M., and Retakh, V.S. (1995). *Noncommutative Vieta Theorem and Symmetric Functions*, preprint Rutgers University, contribution to the Gelfand Seminar Volume.

Gilmore, R. (1974). *Lie Groups, Lie Algebras, and Some of Their Applications*, J.Wiley, New York.

Goldberg, S. I. (1962). *Curvature and Homology*, Dover, New York.

Jackson, J.D. (1975). *Classical Electrodynamics*, 2nd. edition, J. Wiley, New York.

Jones, V.F.R. (1991). *Subfactors and Knots*, AMS, Providence, Rhode Island.

Kobayashi, S. and Nomizu, K. (1963). *Foundations of Differential Geometry*, 2 Volumes, Interscience, New York.

Kirillov, A. (1974). *Élements de la Théorie des Représentations*, MIR, Moscow.

Lee, T.D. (1981). *Particle Physics and Introduction to Field Theory*, Harwood, New York.

Littlewood, D.E. (1950). *The theory of group characters*, Clarendon Press, Oxford.

MacDonald, I.G. (1979). *Symmetric functions and the Hall polynomials*, Oxford Math.Monographs, Oxford University Press.

Nakahara, M. (1990). *Geometry, Topology and Physics*, Institute of Physics Publishing, Bristol and Philadelphia.

Vilenkin, N.Ya. (1969). *Fonctions Spéciales et Théorie de la Représentation des Groupes*, Dunod, Paris.

Zumino, B. (1985). in DeWitt, B.S. and Stora, R. (eds), *Relativity, Groups and Topology* II, vol. 3, North-Holland, Amsterdam.